\documentclass[twocolumn, showpacs, superscriptaddress, a4paper]{revtex4}%
\usepackage{amsfonts}
\usepackage{amsmath}
\usepackage{amssymb}
\usepackage{dcolumn}
\usepackage{bm}
\usepackage{graphicx}
\usepackage{geometry}%
\providecommand{\U}[1]{\protect\rule{.1in}{.1in}}
\begin{document}
\title{Quantum Walks of Two Interacting Anyons in 1D Optical Lattices}
\author{Limin Wang}
\affiliation{Institute of Theoretical Physics, Shanxi University, Taiyuan 030006, P. R. China}
\author{Li Wang}
\affiliation{Institute of Theoretical Physics, Shanxi University, Taiyuan 030006, P. R. China}
\author{Yunbo Zhang}
\email{ybzhang@sxu.edu.cn}
\affiliation{Institute of Theoretical Physics, Shanxi University, Taiyuan 030006, P. R. China}
\date{\today}

\begin{abstract}
We investigate continuous-time quantum walks of two
indistinguishable anyons in one-dimensional lattices with both
on-site and nearest-neighbor interactions based
on the fractional Jordan-Wigner transformation. It is shown that
the two-body correlations in position space are symmetric about
the initial sites of two quantum walkers in the Bose limit
($\chi=0$ ) and Fermi limit ( $\chi=1$), while in momentum space
this happens only in the Bose limit. An interesting asymmetry arises
in the correlation for most cases with the statistical
parameter $\chi$ varying in between. It turns out that the origin
of this asymmetry comes from the fractional statistics that anyons obey.
On the other hand, the two-body correlations of hard-core anyons in position space
show uniform behaviors from anti-bunching to co-walking regardless
of the statistical parameter. The momentum correlations in the
case of strong interaction undergo a smooth process of two stripes smoothly
merging into a single one, i.e. the evolution of fermions into hard-core bosons.

\end{abstract}

\maketitle

\bigskip

\section{INTRODUCTION}

Generally, according to quantum statistical behavior, identical
particles are classified as either bosons, any number of which can
occupy one single-quantum state, or fermions, which occupy a quantum
state exclusively. The exchange of two fermions leads to a phase
factor $-1$ in the total wave function due to the Pauli principle,
whereas the wave function of two bosons remains the same. More than
30 years ago,  a natural generalization was proposed that in
two-dimensional systems there exists a third fundamental category of
identical particles, anyons, which satisfy fractional statistics.
According to the proposal, the overall wave function will acquire a
fractional phase $e^{-i\chi\pi}$ ($0<\chi<1$) when two identical
anyons exchange their positions. From then on, anyon has become a
very important concept in condensed matter physics and has ever been
successfully used in the understanding of fractional quantum Hall
effect (FQHE) \cite{Wilczek,Halperin,Haldane}. All these years, the
research works on anyons remained restricted in the two-dimensional
world \cite{Canright,Wilczek1990} until Haldane put forward the
concept of fractional statistics into arbitrary dimensions
\cite{Haldane}. Nowadays, ultracold atom systems supply as a
versatile toolbox in the field of condensed matter physics and was
proposed to be a wonderful candidate to realize fractional
statistics. Particularly, a proposal to realize anyons in
one-dimensional (1D) optical lattices has already been put forward
in Ref. \cite{Keilmann}. This opens the way to investigate the
exotic properties of anyons in one-dimensional optical lattices both
theoretically and experimentally.

By far, most of the theoretical works
\cite{Batchelorprb,XWGuan,Hao,Hao.Y} on anyon gases in
one-dimensional focus on the ground state properties, like energy,
density profiles, momentum distribution, occupation distribution and
occupations of the lowest natural orbital for different statistical
parameters. Little attention has been paid to anyon properties in
other aspects. In this work, we will study the quantum walks (QWs)
of two identical anyons in one-dimensional optical lattices. Quantum
walks \cite{Aharonov} as the quantum analogs of classical random
walks is a very interesting topic deserving intensive
investigations. It forms the basis of quantum efficient algorithms
and provides a universal platform for quantum computations research.
Compared with classical random walks, quantum walks display several
intriguing non-classical features, such as superpositions and
interference features, which have potential applications in
universal quantum computation \cite{Childs2009,Childs2013} and
detection of topological states \cite{Kitagawa,Kraus} and bound
states \cite{Fukuhara}. What's more, besides single-particle quantum
walks \cite{Karski}, multi-particle quantum walks has attracted more
and more attentions. By employing non-classical correlations
\cite{Benedetti}, multi-particle quantum walks bring new benefits to
practical quantum technologies. Among these, the two-body quantum
walks are of special interest, which have been demonstrated with
both non-interacting photons in linear waveguide arrays
\cite{Hillery,Sansoni,Meinecke} and  interacting photons in
nonlinear waveguide arrays \cite{Solntsev}. And the coexistence of
free and bound states has been explicitly observed through the
quantum walks of two atomic spin-impurities in one-dimensional (1D)
optical lattice(OL) \cite{Fukuhara}. Very recently, there is a
theoretical work investigating two-body quantum walks
\cite{XizhouQin} of bosons, fermions and hard-core bosons.
Specifically, in this paper we shall study the quantum walks of two
identical anyons confined in one-dimensional optical lattices with
both on-site and nearest-neighbor interactions. By using generalized
Jordan-Wigner transformations, we first introduce a mapping from
anyons to bosons or fermions, and then calculate the two-body
correlations in both position and momentum spaces. One thing need to
mention here is that along the line of the excellent experimental
work on Mott insulator \cite{Greiner}, anyons can be simulated using
bosons \cite{Keilmann} with occupation-dependent hopping amplitudes,
which can be realized by assisted Raman tunneling.

The paper is organized as follows. In Sec. II, we introduce the anyon lattice
model with both on-site and nearest-neighbor interactions and construct the
Hilbert space for anyon walkers. The two-body correlations in both position
and momentum spaces are calculated for different evolution time, interaction
strength and statistical parameter and the results are shown in Sec. III.
In Sec. IV, we turn to an alternative definition of the anyon
and study the correlation property of the so-called hard-core anyons.
A brief summary is given in Sec. V.

\section{MODEL AND METHOD\label{secii}}

We consider quantum walks of two indistinguishable anyons in a 1D optical lattice described by
the Hamiltonian with periodic boundary condition
\begin{align}
H^{a}=\sum_{l=-L}^{L} \left[ -J \left(  a_{l}^{\dagger}a_{l+1}+h.c\right)
\right. \nonumber\\
\left.  +\frac{U}{2}n_{l}\left(  n_{l}-1\right)  +V n_{l}n_{l+1} \right]  .
\label{H}%
\end{align}
Here $a_{l}^{\dagger}\left(  a_{l}\right)  $ creates (annihilates)
an anyon on the $l$-th site, $n_{l}=a_{l}^{\dagger}a_{l}$ is the
particle number, $J$ is the hopping between neighboring sites, $U$
and $V$ describe the on-site and nearest-neighbor interactions,
respectively. We consider the initial condition being that two
particles are localized at adjacent lattice sites and the dynamics
on $2L+1$ lattice sites represents a typical QW problem of
continuous time.

The commutation relations \cite{Kundu,Batchelor,Ovidiu} (CRs) for the anyonic
operators read as%
\begin{align}
a_{l}a_{k}^{\dagger}  &  =e^{-i\chi\pi\epsilon\left(  l-k\right)  }%
a_{k}^{\dagger}a_{l}+\delta_{lk},\nonumber\\
a_{l}a_{k}  &  =e^{i\chi\pi\epsilon\left(  l-k\right)  }a_{k}a_{l},\nonumber\\
a_{l}^{\dagger}a_{k}^{\dagger}  &  =e^{i\chi\pi\epsilon\left(  l-k\right)
}a_{k}^{\dagger}a_{l}^{\dagger}. \label{crs}%
\end{align}
The sign function $\epsilon\left(  x\right)  $ gives $-1,0,$ or $1$
depending on whether $x$ is negative, zero, or positive. To describe
the fractional statistics of the anyon, it is evident that the range
of parameter $\chi$ is sufficient to be restricted in the interval
$\chi\in\left[  0,1\right]  $. In the original work \cite{Kundu}
introducing anyonic system model, the anyonic fields were usually
realized in terms of the bosonic fields. This assures that the
anyonic system reduces to the bosonic system naturally in the limit
$\chi=0$. The same idea, called the anyon-boson mapping, will be
used below. However, anyons with statistics $\chi=1$ are
pseudo-fermions: being fermions off-site, they are nevertheless
bosons on-site. An alternative realization of anyonic fields in
terms of the fermionic fields was proposed to eliminate the
difficulties for zero-range interaction \cite{Girardeau}. This
fermionic representation with appropriate modification of the
statistical parameter $\chi$ makes it possible to describe only the
infinite repulsive limit, i.e. the hard-core interaction, which will
be adopted to study the correlation of hard core anyons in Sec. IV.

In order to study the correlation property of anyons, we resort to
an exact mapping between anyons and bosons in 1D. Let us introduce
the fractional version of a Jordan--Wigner transformation
\cite{Keilmann}
\begin{align}
a_{l}  &  =b_{l}\exp\left(  -i\chi\pi\sum_{i=-L}^{l-1}n_{i}\right)
,\nonumber\\
a_{l}^{\dagger}  &  =\exp\left(i\chi\pi\sum_{i=-L}^{l-1}n_{i}\right)
b_{l}^{\dagger}, \label{j-w}%
\end{align}
with $n_{l}=a_{l}^{\dagger}a_{l}=b_{l}^{\dagger}b_{l}$ the number operator for
both particle types. Starting from the
bosonic CRs, i.e. $\left[  b_{l},b_{k}\right]  =\left[  b_{l}^{\dagger}%
,b_{k}^{\dagger}\right]  =0$, and $\left[
b_{l},b_{k}^{\dagger}\right] =\delta_{lk}$, we can validate that the
mapped operators $a_{l}$ actually obey the anyonic commutation
relations as introduced in Eq. (\ref{crs}). This mapping elucidates
that anyons in 1D are indeed non-local quasi-particles, made of
bosons with an attached string operator.

Our final goal is to propose a realistic method for studying the dynamics of an
interacting gas of anyons in 1D OLs. Hence, by means of the anyon--boson
mapping (\ref{j-w}), and considering the periodic boundary condition of anyons
in 1D lattices \cite{Rigol},%
\begin{align}
a_{L}^{\dagger}a_{-L}  &  =\exp\left(
i\chi\pi\sum_{i=-L}^{L-1}n_{i}\right)
b_{L}^{\dagger}b_{-L},\nonumber\\
a_{-L}^{\dagger}a_{L}  &  =b_{-L}^{\dagger}b_{L}\exp\left(
-i\chi\pi\sum _{i=-L}^{L-1}n_{i}\right)  ,
\end{align}
the Hamiltonian $H^{a}$ can be rewritten in terms of bosonic operators,%
\begin{align}
H^{b} =\text{ }  &  -J\sum_{l=-L}^{L-1}\left(  b_{l}^{\dagger}\exp\left(
-i\chi\pi n_{l}\right)  b_{l+1}+h.c\right) \nonumber\\
&  -J\left(  \exp\left(  i\chi\pi\sum_{i=-L}^{L-1}n_{i}\right)
b_{L}^{\dagger
}b_{-L}+h.c\right) \nonumber\\
&  +\frac{U}{2}\sum_{l=-L}^{L}n_{l}\left(  n_{l}-1\right)  +V\sum_{l=-L}%
^{L}n_{l}n_{l+1}. \label{hamib}%
\end{align}
The mapped, bosonic Hamiltonian thus describes bosons with an
occupation-dependent amplitude $J\exp\left(  -i\chi\pi n_{l}\right)
$ for hopping processes between adjacent sites ($l, l+1$) except on
the boundary ($-L$ and $L$). If the target site $l$ is unoccupied,
the hopping amplitude is merely $J$. If it is occupied by one boson,
the amplitude reads $J\exp\left(  -i\chi\pi\right)  $, and so on. We
underline that the non-local mapping between anyons and bosons, Eq.
(\ref{j-w}), leads luckily to a purely local, and thus viable
Hamiltonian. As expected from anyons, the reflection parity symmetry
is broken at the level of the commutation relations (\ref{crs}). The
fractional Jordan--Wigner transformation (\ref{j-w}) transfers this
asymmetry also to the bosonic case: the resulting Hamiltonian
(\ref{hamib}) features a phase factor acting only on the target site
$l$ and thus violates parity.

We now discuss the Hilbert space involved by the QWs of two particles. Since
$\left[  N,H\right]  =0$, the total particle number $N$ is conserved and the
system will evolve in the two-particle Hilbert space. For two anyons, their
Hilbert space can be spanned by basis, $B_{a}^{\left(  2\right)  }=B_{b}^{\left(  2\right)  }=\left\{
\left\vert l_{1}l_{2}\right\rangle =\left(  1+\delta_{l_{1}l_{2}}\right)
^{-\frac{1}{2}}b_{l_{1}}^{\dagger}b_{l_{2}}^{\dagger}\left\vert \mathbf{0}%
\right\rangle ,-L\leq l_{1}\leq l_{2}\leq L\right\}  $ because of the exact
mapping between anyons and bosons.
Here, $\left\vert \mathbf{0}\right\rangle $ denotes the vacuum state.
Given $B_{b}^{2}$, it is easy to construct the
Hamiltonian matrix $H^{(2)}$ in two-particle sector. In units of $\hbar=1$,
the time evolution of an arbitrary state obeys%
\begin{equation}
i\frac{d}{dt}\left\vert \psi\left(  t\right)  \right\rangle =H^{\left(
2\right)  }\left\vert \psi\left(  t\right)  \right\rangle , \label{hs}%
\end{equation}
with $\left\vert \psi\left(  t\right)  \right\rangle =\sum_{l_{1}\leq l_{2}%
}C_{l_{1},l_{2}}\left(  t\right)  \left\vert l_{1}l_{2}\right\rangle $ for
anyons. Below we will study the continuous-time QWs of two anyons
induced by the statistics parameter $\chi$ starting from an initial state $\left\vert
\psi_{initial}\right\rangle =a_{0}^{\dagger}a_{1}^{\dagger}\left\vert
\mathbf{0}\right\rangle $, where the two anyons are prepared in two
adjacent sites $0$ and $1$. In order
to explore the correlation between two quantum walkers, we calculate the
two-body correlation in position space,%
\begin{equation}
\Gamma_{qr}\left(  t\right)  =\left\langle \psi\left(  t\right)  \right\vert
a_{q}^{\dagger}a_{r}^{\dagger}a_{r}a_{q}\left\vert \psi\left(  t\right)
\right\rangle , \label{dlzb}%
\end{equation}
and that in momentum space,%
\begin{equation}
\Gamma_{\alpha\beta}\left(  t\right)  =\left\langle \psi\left(  t\right)
\right\vert a_{\alpha}^{\dagger}a_{\beta}^{\dagger}a_{\beta}a_{\alpha
}\left\vert \psi\left(  t\right)  \right\rangle. \label{dldl}%
\end{equation}
The operators in momentum space are defined by the discrete Fourier
transformation
\begin{eqnarray}
a_{\alpha}^{\dagger
}&=&\frac{1}{\sqrt{2L+1}}\sum_{l=-L}^{L}e^{-ip_{\alpha}l}a_{l}^{\dagger}, \notag \\
a_{\alpha}&=&\frac{1}{\sqrt{2L+1}}\sum_{l=-L}^{L}e^{ip_{\alpha}l}a_{l},
\end{eqnarray}
with the quasi-momentum $p_{\alpha}=\frac{2\pi\alpha}{2L+1}$ and
the indices
$q,r,\alpha,\beta=\left(  -L,\cdots,0,\cdots,L\right) $. The correlation matrix
$\Gamma_{qr}\left(  t\right)  $ represents the probability of detecting one
particle at site $q$ and its twin particle at site $r$, which is
calculated after different evolution times $Jt$ for different values of the
interaction strengths $V$ or $U$. Similar probability interpretation is
imposed on $\Gamma_{\alpha\beta}\left(  t\right)$, however, in the
momentum space. At all stages the particles are far from
the lattice boundaries - the signal due to the boundary condition will
be discussed elsewhere. The correlations in the following figures are
rescaled by their maximum values such that $\Gamma_{qr}\left(  t\right)
/\Gamma_{qr}^{\max}\left(  t\right)  $ and $\Gamma_{\alpha\beta}\left(  t\right)
/\Gamma_{\alpha\beta}^{\max}\left(  t\right)  $) are shown.

\begin{figure}[t]
\includegraphics[width=0.5\textwidth]{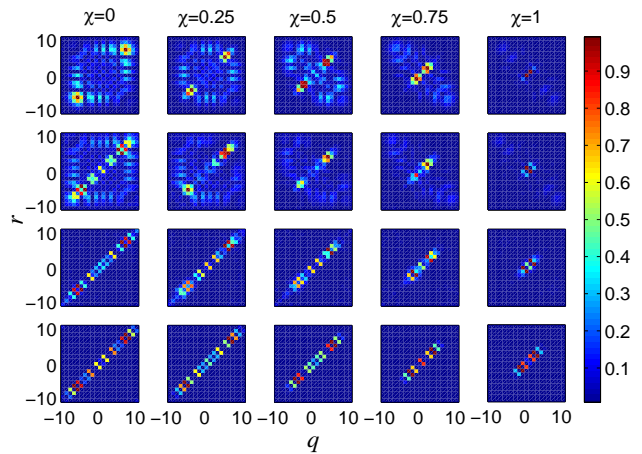} \caption{(Color online)
Two-body correlations of anyonic quantum walkers in position space.
The nearest-neighbor interaction strength $\left\vert V/J\right\vert
=0,1,4$ and $80$ from top to bottom and the on-site interaction
strength $U/J=0$. Here we only show the instantaneous correlations
before colliding with the boundaries $L=\pm10$. The corresponding
evolution times are $Jt=4,4.5,7.5$ and $110$
from top to bottom. For each row, $\chi=0,0.25,0.5,0.75,1$ (from left to right ).}%
\label{FIG.1.}%
\end{figure}

\begin{figure}[t]
\includegraphics[width=0.5\textwidth]{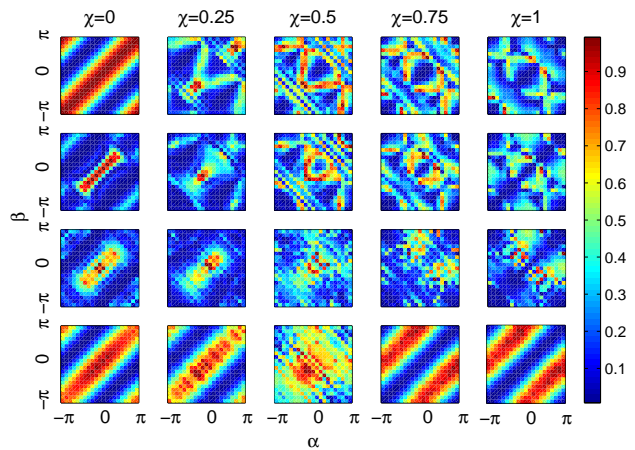} \caption{(Color online)
Two-body correlations of anyonic quantum walkers in momentum space with the same
parameters as in Fig. 1. Note that spatial asymmetry in the bottom panel for $\chi=1$
persists for very strong interaction.}%
\label{FIG.2.}%
\end{figure}

\section{Correlation of two interacting anyons}

We first investigate the two-body correlations in both position and
momentum spaces in absence of the on-site interaction, i.e. $U/J=0$.
To be more specific, we let the nearest-neighbor interaction be
attractive $V<0$ in the whole paper. Following the procedure
described in Sec. II, the two-body correlations in position and
momentum spaces can be exactly obtained and are shown in Fig.
\ref{FIG.1.} and Fig. \ref{FIG.2.}. The effects induced by the
statistical parameter $\chi$ provide clear insights into the exotic
behavior of anyonic two-body QWs.

As shown in Fig. \ref{FIG.1.},  the first column is corresponding to
the Bose limit ($\chi=0$), where the obvious bunching behavior
appears in the two-body correlations. This is exactly what has been
shown in \cite{XizhouQin}. Two bunching points along the principal
diagonal line get closer and closer with increasing $\chi$. The last
column is for the Fermi limit ($\chi=1$), where the expected
anti-bunching behavior is not seen. This is due to that the Fermi
limit here corresponds to pseudo-fermions instead of real fermion
walkers. Pseudo-fermions in our system behave as follows: being
bosons on-site, they tend to stay together in any one of the initial
sites when there is no interaction; being fermions off-site, they
start to occupy adjacent lattice sites for finite interaction $V$
and stick together when co-walking in opposite directions with equal
probability. For strong interaction $V$, correlations due to the
independent walking become invisible and the real fermion behavior
returns \cite{XizhouQin} (see the lower-right panel in Fig.
\ref{FIG.1.}).

Furthermore, we find that the correlations in the Bose limit ($\chi=0$) and in the Fermi
limit ($\chi=1$) are both symmetric about the initial positions of two anyons.
However, in the presence of a finite strength of nearest-neighbor interaction
(the second and third rows in Fig. \ref{FIG.1.}),
the correlations are found to be asymmetric once the statistical parameter $\chi$ deviates
from the two limits. This can be understood as a pure effect of the statistical parameter $\chi$.
With increasing $\chi$ the occupation-dependent statistical
factor in (\ref{hamib}) becomes more and more important: the tunneling processes
connecting sites with different occupations will contribute different values.
We see difference when $n_l=1$: the hopping term to the left will gain a phase factor
$e^{-i\chi \pi}$, the term to the right, on the other hand, acquires $e^{i\chi \pi}$.
In the kinetic part of the Hamiltonian this incoherent superpositions is amplified by an
increasing $\chi$ and induces the asymmetry of the correlation.
The attractive interaction between nearest-neighboring sites forces the two anyons
co-walking together indicated as the significant correlations at the two secondary diagonal lines.
Stronger interaction dominates the dynamics and the bunching and co-walking behavior revives
as shown in the last row in Fig. \ref{FIG.1.}.

The two-body correlations in momentum space are shown in Fig.
\ref{FIG.2.}. The first column of Fig. \ref{FIG.2.} recovers again
the behavior of bosonic walkers in Ref. \cite{XizhouQin}, while the
last column fails to converge to the fermi limit due to the
pseudo-fermion nature of the anyons when $\chi=1$. Compared with the
position space in Fig. \ref{FIG.1.},  the effects of statistical
parameter $\chi$ on the asymmetry of the two-body correlations are
much clearer in momentum space. In Fig. \ref{FIG.2.}, we find that
once the statistical parameter $\chi$ deviates from the Bose limit
($\chi=0$), the interesting asymmetry emerges immediately in the
two-body correlations - we even do not need finite interaction
strength. It is thus concluded that the asymmetry originates from
the exotic statistical properties of the anyons. For strong enough
interaction, the spatial correlations of two anyonic walkers
degenerate into the real fermion walkers successfully, while in
momentum space we find clear signature of crossover from bunching
behavior at $\chi=0$ to anti-bunching behavior at $\chi=1$ but the
asymmetry persists for very strong interaction as can been seen in
the last row of Fig. \ref{FIG.2.}.

\begin{figure}[t]
\includegraphics[width=0.5\textwidth]{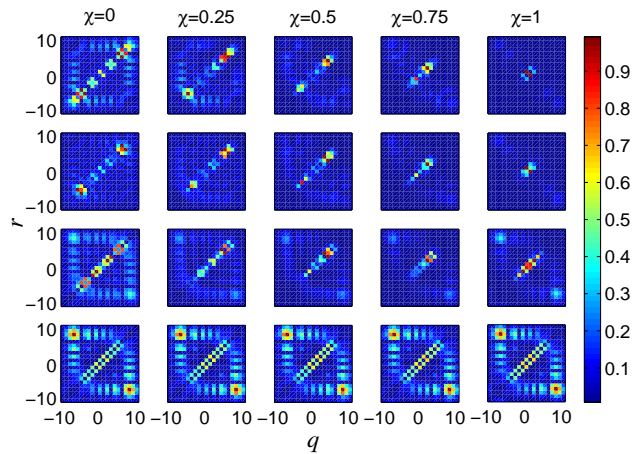} \caption{(Color online)
Two-body correlations of anyonic quantum walkers in position space.
The on-site interactions are $U/J=0,1,4$ and $80$ (from top to
bottom) with fixed $\left\vert V/J\right\vert =1$ and the
corresponding
evolution time is given by $Jt=4.5$. }%
\label{FIG.3.}%
\end{figure}

Next we introduce the on-site interaction $U$ and see how it will
compete with the nearest-neighbor interaction $V$. In
Fig. \ref{FIG.3.}, $V$ is fixed to $|V/J|=1$ and we vary $U$ in a
range from $0$ to $80$ in order to shed a light on the different
roles played by the two interaction terms in the Hamiltonian. The
first column of Fig. \ref{FIG.3.} shows the Bose limit and we find
that as on-site interaction $U$ increases, the two-body
correlations evolve from a bunching behavior (bosonic walkers) to an
anti-bunching behavior (hard-core bosonic walkers). Similar shrinking
tendency is observed with increasing statistical parameter $\chi$. In
the Fermi limit, we see that the behavior of the
two-body correlations evolves from pseudo-fermions to real
fermions. With large enough $U$, the two-body
correlations of the hard-core bosons, hard-core anyons and real fermions are almost
identical in the last row of Fig. \ref{FIG.3.}.

\begin{figure}[t]
\includegraphics[width=0.5\textwidth]{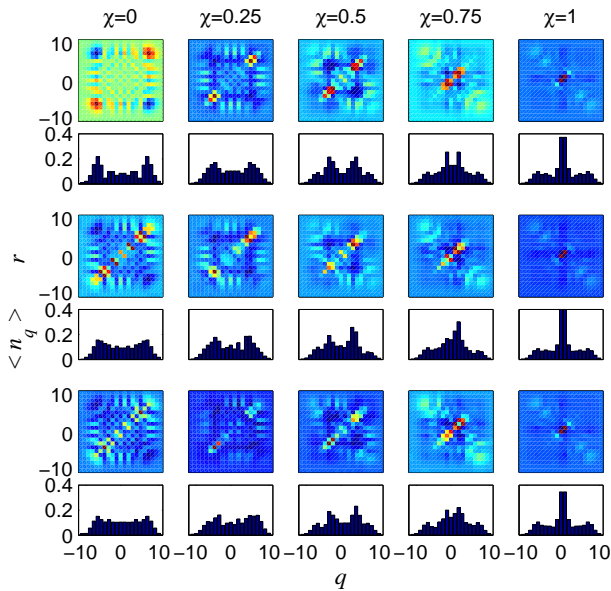} \caption{(Color online)
Two-body correlation fluctuations and the density distributions in position
space. The on-site and nearest-neighbor interaction strengths are
$(U/J, \left\vert V/J \right\vert )=(0,0), (0,1)$ and $(1,0)$ from top to bottom.
The evolution time is given by $Jt=4$. }
\label{FIG.4.}
\end{figure}

It is also interesting to show the effect of the statistics
parameter $\chi$ on the correlation fluctuation \cite{Lahini}
defined as
$\Gamma_{qr}^{F}$ $\left(  t\right)  =\left\langle a_{q}^{\dagger}%
a_{r}^{\dagger}a_{r}a_{q}\right\rangle -\frac{1}{2}\left\langle a_{q}%
^{\dagger}a_{q}\right\rangle \cdot\left\langle a_{r}^{\dagger}a_{r}%
\right\rangle =\Gamma_{qr}-\frac{1}{2}\left\langle
n_{q}\right\rangle \cdot \left\langle n_{r}\right\rangle$ and the
particle density $\left\langle n_{q}\right\rangle=\left\langle
a_{q}^{\dagger}a_{q}\right\rangle $. In Fig. \ref{FIG.4.}, we find
that the statistical parameter $\chi$ breaks the symmetry in both
the correlation fluctuation and the particle density for finite $U$
or $V$. The statistics parameter plays an essential role in the
symmetry of the density distribution, in contrast with the case of
hard-core anyons \cite{Hao.Y} where the statistical factor makes no
difference. The asymmetry here again arises from the
statistics-dependent hopping term in Hamiltonian (\ref{hamib}).

\section{Correlation of two hard-core anyons}

In view of the above-mentioned facts, i.e. two-body correlation
calculated through the mapping between anyons and bosons fails to
converge to the Fermi limit due to the pseudo-fermion nature of the
anyons for $\chi=1$, we turn to an alternative realization of
anyonic fields in terms of the fermionic fields \cite{Girardeau}.
The mapping now is between anyons and fermions, which guarantees
that anyons would return to fermions in the limit $\kappa=0$. Here
we introduce another parameter $\kappa$ to describe the statistical
property of this newly defined anyon. The anyonic operators now
satisfy CRs \cite{Hao,Hao.Y,Girardeau,Batchelor2008}
\begin{align}
a_{l}a_{k}^{\dagger}  &  =\delta_{lk}-e^{-i\kappa\pi\epsilon\left(
l-k\right)  }a_{k}^{\dagger}a_{l},\nonumber\\
a_{l}a_{k}  &  =-e^{i\kappa\pi\epsilon\left(  l-k\right)  }a_{k}%
a_{l},\nonumber\\
a_{l}^{\dagger}a_{k}^{\dagger}  &  =-e^{i\kappa\pi\epsilon\left(
l-k\right)
}a_{k}^{\dagger}a_{l}^{\dagger}. \label{crs2}%
\end{align}
Then the exclusion principle $a_{l}^{\dagger2}=a_{l}^{2}=0$ and
$\left\{ a_{l},a_{l}^{\dagger}\right\} =1$ follow from
$\epsilon\left(  x\right)=0 $. They are the commutation relations
which are frequently used to describe hard-core condition
\cite{Rigol}. Anyons with these properties are called hard-core
anyons (HCAs) \cite{Batchelor2008}. The CRs (\ref{crs2}) connect
real fermions and hard-core bosons when $\kappa=0$ and $1$, just as
CRs (\ref{crs}) link ordinary bosons and pseudo-fermions when
$\chi=0$ and $1$. In this section we essentially investigate the
quantum walks of two HCAs in 1D optical lattice. The Hamiltonian of
hard-core anyons reads
\begin{equation}
H^{HCA}=-J\sum_{l=-L}^{L}\left(  a_{l}^{\dagger}a_{l+1}+h.c\right)
+V\sum_{l=-L}^{L}n_{l}n_{l+1}. \label{hamihca}%
\end{equation}
This model can also be exactly solved through a generalized
Jordan-Wigner transformation  mapping the hard-core anyons to
spinless fermions \cite{Girardeau},
\begin{align}
a_{l}  &  =f_{l}\exp\left(  -i\kappa\pi\sum_{i=-L}^{l-1}n_{i}\right)
,\nonumber\\
a_{l}^{\dagger}  &
=\exp\left(i\kappa\pi\sum_{i=-L}^{l-1}n_{i}\right)
f_{l}^{\dagger}, \label{j-w2}%
\end{align}
where $f_{l}^{\dagger}$ and $f_{l}$ are creation and annihilation
operators for spinless fermions,
$n_{l}=a_{l}^{\dagger}a_{l}=f_{l}^{\dagger}f_{l}$ is particle number
operator for both particle types. By means of this mapping, the
hard-core anyonic Hamiltonian (\ref{hamihca}) can be described by
fermion operators,
\begin{align}
H^{f} =\text{ }  &  -J\sum_{l=-L}^{L-1}\left(  f_{l}^{\dagger}f_{l+1}%
+h.c\right)  \nonumber \\
&  -J\left(  \exp\left(  i\kappa\pi\sum_{i=-L}^{L-1}n_{i}\right)
f_{L}^{\dagger}f_{-L}+h.c\right) \nonumber \\
&  +V\sum_{l=-L}^{L}n_{l}n_{l+1} \label{splf},
\end{align}
on which the periodic boundary condition is imposed. We notice that
a big difference here is that the statistical factor appears only on
the boundary due to the hard core constraint. The corresponding
Hilbert space of the two quantum walkers can be constructed as
$B_{HCA}^{(2)}=B_{f}^{(2)}=\left\{  \left\vert
l_{1}l_{2}\right\rangle
=f_{l_{1}}^{\dagger}f_{l_{2}}^{\dagger}\left\vert
\mathbf{0}\right\rangle ,-L\leq l_{1}<l_{2}\leq L\right\}  $.
Again the initial state is chosen as $\left\vert \psi_{initial}%
\right\rangle =a_{0}^{\dagger}a_{1}^{\dagger}\left\vert \mathbf{0}%
\right\rangle $.

\begin{figure}[t]
\includegraphics[width=0.5\textwidth]{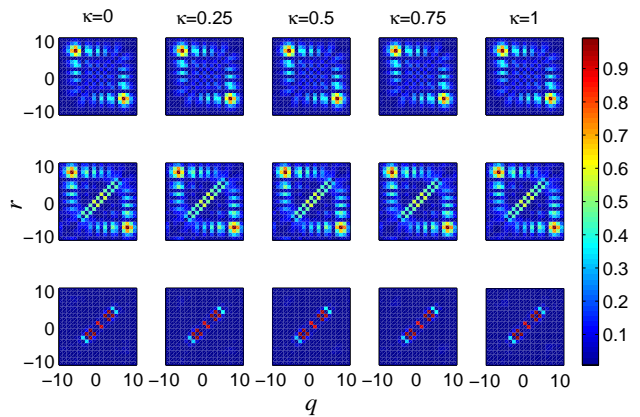} \caption{(Color online)
Two-body correlations in position space for HCAs. The nearest-neighbor interaction
strength $\left\vert V/J\right\vert =0,1,$ and $4$ from top to bottom
and the corresponding evolution times are given by
$Jt=4,4.5,$ and $7.5$, respectively. In each row, $\kappa
=0,0.25,0.5,0.75,1$ (from left to right).}%
\label{FIG.5.}%
\end{figure}

\begin{figure}[t]
\includegraphics[width=0.5\textwidth]{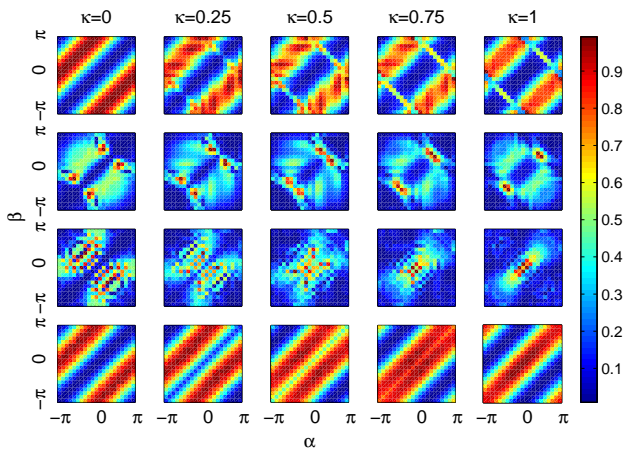} \caption{(Color online)
Two-body correlations in momentum space for HCAs.
The nearest-neighbor interactions
are $\left\vert V/J\right\vert =0,1,4$ and $80$ from top to bottom.
And the corresponding evolution times are
given by $Jt=4,4.5,7.5,$ and $110,$ respectively.
The figures in the first and last column
are exactly the same as  those in \cite{XizhouQin} .}%
\label{FIG.6.}%
\end{figure}

By the similar procedure described in Sec. \ref{secii}, we calculate
the two-body correlations both in position and momentum spaces for
HCAs. In position space, as shown in Fig. \ref{FIG.5.}, we find that
the two-body correlations echo no signal of the statistical
parameter $\kappa$ and are all the same as those for spinless
fermions \cite{XizhouQin}. In each row, from left to right, the
pictures are almost the same to naked eye though slight change
happens according to the numerical data. This is because the
Hamiltonian Eq. (\ref{splf}) is the same as spinless fermions except
the boundary terms. With increasing $V$ the anyons show uniform
transition from anti-bunching to co-walking regardless of the
statistical factor. The diagonal elements of the correlation vanish
due to the hard core condition, which can be seen from
\begin{eqnarray}
\Gamma_{qr}  &  =&\left\langle \psi^{HCA}\left(  t\right)  \right\vert
a_{q}^{\dagger}a_{r}^{\dagger}a_{r}a_{q}\left\vert \psi^{HCA}\left(  t\right)
\right\rangle  \notag \\
&=&\left\langle \psi^{HCA}\left(  t\right)  \right\vert
n_{q} n_{r}\left\vert \psi
^{HCA}\left(  t\right)  \right\rangle,
\end{eqnarray}
where the  statistical factors cancel with each other automatically. However,
this is not the case for two-body correlations in momentum
space. To obtain the correlation $\Gamma_{\alpha\beta}\left(
t\right)  =\left\langle
\psi^{HCA}\left(  t\right)  \right\vert a_{\alpha}^{\dagger}a_{\beta}%
^{\dagger}a_{\beta}a_{\alpha}\left\vert \psi^{HCA}\left(  t\right)
\right\rangle $, one has to calculate terms like $ \left\langle
\psi^{HCA}\left(  t\right)  \right\vert
a_{p}^{\dagger}a_{q}^{\dagger}a_{r}a_{s}\left\vert \psi^{HCA}\left(
t\right) \right\rangle$, where the four indices are generally
different. Thereby, the  statistical factors introduced by the
mapping Eq. (\ref{j-w2}) will stay. The two-body correlations in
momentum space carry more informations as shown in Fig.
\ref{FIG.6.}.

In Fig. \ref{FIG.6.}, the first and last columns correspond
precisely to the Fermi limit $\kappa=0$ for real fermions and the
Bose limit $\kappa=1$ for hard-core bosons, respectively, which are
the main topic of Ref. \cite{XizhouQin}. Once the statistical factor
$\kappa$ deviates from the two limits, the asymmetry of correlation
reappears and in between there again exists a crossover from an
anti-bunching behavior (fermionic walkers) to a bunching behavior
(hard-core bosonic walkers) with increasing $\kappa$. Especially,
for strong interaction, the last row shows how the two stripes in
the momentum correlation smoothly merge into a single one. This
provides a way to probe different statistics from the observations
of the correlations in momentum space, which is viable in ultracold
atom laboratories nowadays.

\section{CONCLUSIONS}

In summary, we have investigated the two-body correlations of 1D
quantum gas of anyons confined in optical lattices with both on-site
and nearest-neighbor interactions using exact numerical method. With
Jordan-Wigner transformation the anyon lattice model is mapped to
bosonic one and thus the Hilbert space of anyons can be constructed
from that of bosons. Then by solving the time-dependent
Schr\"{o}dinger equation we obtain the wave function in arbitrary
evolution time. Numerical results show that the anyonic two-body
correlations in position space exhibit distinct properties from the
bosons and fermions. In the Bose and Fermi limits the correlations
are symmetric about the initial positions of walkers in position
space. The variation in statistic parameter $\chi$ drives the system
from Bose statistics to Fermi statistics and the fractional
statistics in between. An interesting asymmetry arises in the
correlations in both position and momentum spaces due to the
fractional statistics of the anyons. The anyon-boson mapping links
the bosons and pseudo-fermions when $\chi=0$ and $1$, respectively,
while for HCAs, the anyon-fermion mapping connects the fermions and
hard-core bosons for $\kappa=0$ and $1$. In this sense we conclude
that anyons realized in the two ways are not simply intermediate
particles between bosons and fermions. In either case, the
correlations only converge to one limit, through which the anyons
are defined. Stronger inter-site interaction is needed in achieving
perfect evolution from bosonic walkers to fermionic walkers with
increasing $\chi$. On the other hand, stronger on-site interaction
will put a hard-core constraint on anyons, the statistics of which
can be distinguished from the correlation in momentum space.

\begin{acknowledgments}
This work is supported by the NSF of China under Grant Nos. 11234008,
11104171, 11404199 and 11474189, the National Basic Research Program of China (973
Program) under Grant No. 2011CB921601, Program for Changjiang Scholars
and Innovative Research Team in University (PCSIRT)(No. IRT13076).
\end{acknowledgments}

\end{document}